\documentclass{ws-ijbc} 
\usepackage{amsmath}\usepackage{calc,color}\usepackage{amssymb}
\usepackage{subfigure}
\usepackage{ws-rotating}     
\usepackage{graphicx}
\usepackage{epstopdf}
\begin{document}

\catchline{}{}{}{}{} 

\markboth{Alonso-Sanz and Adamatzky}{ACTIN AUTOMATA WITH MEMORY}

\title{ ACTIN AUTOMATA WITH MEMORY}
\author{Ram\'on Alonso-Sanz}
\address{Technical University of Madrid, ETSIA\,(Estadistica,GSC)\\
 ramon.alonso@upm.es}
\author{Andy Adamatzky}
\address{University of the West of England, Bristol, UK. \\
 andrew.adamatzky@uwe.ac.uk}
\maketitle
 
\begin{abstract}
Actin is a globular protein which forms long polar filaments in eukaryotic. The actin filaments play roles of cytoskeleton, motility units , information processing and learning.  We model actin filament as a double chain of finite state machines, nodes, which take states `0' and `1'. The states are abstractions of absence and presence of a sub-threshold charge on an actin units corresponding to the nodes. All nodes update their state in parallel in discrete time. A node updates its current state depending on states of two closest neighbours in the node chain and two closest neighbours in the complementary chain. Previous models of actin automata considered momentary state transitions of nodes. We enrich the actin automata model by assuming that states of nodes depends not only on the current states of neighbouring node but also on their past states. Thus, we assess the effect of memory of past states on the dynamics of acting automata. We demonstrate in computational experiments that memory slows down propagation of perturbations, decrease entropy of space-time patterns generated, transforms travelling localisations to stationary oscillators, and stationary oscillations to still patterns.  
\end{abstract}


\section{Introduction}

Actin is a globular protein presented as intracellular, cytoskeleton, filaments  in eukariotic cells  from unicellular organisms to plants and animals.   Actin filaments, together with  tubule microtubule filaments, are not only key components responsible for coordinated motility of cells but are `nervous system' of the cells. The actin and tubule filaments process information and implement learning~\cite{hameroff1988coherence, rasmussen1990computational, ludin1993neuronal, conrad1996cross, tuszynski1998dielectric, priel2006dendritic, debanne2004information, priel2010neural, jaeken2007new}.  Disfunction in actin assembly or the actin association with other intracellular components leads to psychiatric and neurological disorders~\cite{van1990major, fiala2002dendritic, persico2006searching, sekino2007role, van2010synapse, kojima2007synaptic}. Therefore, by uncovering `mechanics' of signal/pertubartion propagation on actin filaments we can, in future, develop novel principles of information processing at sub-cellular level, and may be even contribute towards nano medicine based treatments of neurological disorders.

Computational models tubulin microtubules has been developed in 1990s and used to demonstrate that computation could be implemented in tubulin protofilaments by classical and quantum means~\cite{rasmussen1990computational, hameroff1992models, janmey, hameroff}. Less attention was paid to actin double helix filaments, despite importance of the actin in learning and information pre-processing as might be hinted by pre-dominant presence of actin networks in synapses~\cite{fifkova1982cytoplasmic, kim1999role, dillon2005actin, cingolani2008actin}.  Previously  we proposed a model of  actin filaments as two chains of one-dimensional binary-state semi-totalistic automaton arrays~\cite{aadamatzky}. We analysed the complete rule space of actin automata using integral characteristics of space-time configurations generated by these rules and compute state transition rules that support travelling and mobile localizations. We found that some properties of actin automata rules may be predicted using Shannon entropy, activity and incoherence of excitation between the polymer chains. We also shown that it is possible to infer whether a given rule supports travelling or stationary localizations by looking at ratios of excited neighbours that are essential for generations of the localizations.


 \section{Automata with memory}

Conventional CA are memoryless\,: i.e., the new
state of a cell depends on the neighborhood configuration solely at the
preceding time step. The standard framework of CA  can be extended by  implementing memory 
 capabilities in cells\,\,\cite{alonso2011discrete,alonso2006elementary,alonso2010One-dimensional}\,: 
$$\sigma^{(T+1)}_{i}=\phi\big(\{{s}^{(T)}_{j}\}\in\mathcal{N}_{i}{\big )},$$
\par\noindent
with $s^{(T)}_j$ being a state function of the series of states of the cell $j$ up to \texttt{T} :
\begin{equation}\label{summary}
{s}^{(T)}_{j}={s}\big(\sigma^{(1)}_{j},\ldots,\sigma^{(T-1)}_{j}, \sigma^{(T)}_{j}\big) 
\end{equation}
\par
 Thus in CA with memory,  while the mappings $\phi$ remain unaltered, historic memory of all past iterations is 
retained by featuring each cell as a summary of its past states. So to say, cells $canalize$ memory to the map $\phi$. 
\par
Memory \eqref{summary} may be implemented as majority memory, i.e., the most frequent (mode) state\,:
\begin{equation}\label{modememory}
 \, s^{(T)}_{i}=mode(\sigma^{(1)}_{i},\ldots,\sigma^{(T)}_{i})
\end{equation}
 with $~s^{(T)}_{i}$=$\sigma^{(T)}_{i}$ in case of a tie: $card\{1\}=card\{0\}$\,.
\par
At variance with the unlimited memory implementation in \eqref{modememory}\,, 
the length of the trailing memory may be limited to the last $\tau$ time-steps. 
\begin{equation}\label{taumodememory}
s^{(T)}_{i}=mode(\sigma^{(T-\tau+1)}_{i},\ldots,\sigma^{(T-2)}_{i},\sigma^{(T-1)}_{i},\sigma^{(T)}_{i})\,.
\end{equation}
\par
The shortest operative trailing length is that of $\tau=3$\,:
\begin{equation}\label{taumodememory-}
s^{(T)}_{i}=mode(\sigma^{(T-2)}_{i},\sigma^{(T-1)}_{i},\sigma^{(T)}_{i})\,.
\end{equation}
with initial assignations\,: $~s^{(1)}_{i}$=$\sigma^{(1)}_{i}$,  $~s^{(2)}_{i}$=$\sigma^{(2)}_{i}$\,.
\par 
Keeping track of the states of the last $\tau$ time-steps  demands $\tau$ extra bits per cell to store
their corresponding state values. To avoid this drawback, past state values can be weighted in such a
way that only the accumulated memory charge needs to be stored.  
Thus, for example, historic memory can be weighted by applying a geometric discounting process
in which the state $\sigma ^{(T-\tau)}_{i}$, obtained $\tau$ time steps
before the last round, is actualized to  $\alpha ^{\tau}\sigma
^{(T-\tau)}_{i}$, $\alpha$ being the \textit{memory factor} lying in the [0,1]
interval. This well known mechanism fully takes into account the last round $(\alpha
^{0}=1)$, and tends to \textit{forget} the older rounds. Thus, 

\begin{equation}\label{eq:omegaeq}
\omega^{(T)}_{i}(\sigma ^{(1)}_{i},\ldots ,\sigma ^{(T)}_{i}) = 
\sigma^{(T)}_{i} + \displaystyle {\sum^{T-1}_{\tau=0}\alpha ^{\tau}\sigma^{(T-\tau)}_{i}} 
= \sigma^{(T)}_{i} + \alpha \omega^{(T-1)}_{i}
\end{equation}

\par
 Consequently, only one  number per cell ($\omega_i$) needs to be stored. This positive property is accompanied by the drawback
that it computes with real numbers, which is not in the realm
of CA, that claims for integer arithmetics. 
\par 
    Every cell will be featured first by the weighted mean ($m$)  of all its past
states, so the memory charge at  time-step $T$ is\,: 
\par

\begin{equation}\label{eq:meq}
m^{(T)}_{i}= {\displaystyle\frac{{{\omega^{(T)}_{i}}} }{\Omega(T) }}\hspace{0.15cm}\,,\hspace{0.25cm}
~~\hbox{with}~~ \Omega{(T)}= {1+ \displaystyle \sum^{T-1}_{t=1}\alpha ^{T-t}}  
\end{equation}
\par
The trait state \textit{s} is obtained by
comparing $m$ to the landmark  0.5  (if $\sigma\in\{0,1\}$), assigning the last state in case 
of an equality to this value, so that\, :\vspace{-.25cm}

\begin{equation}\label{rounding}
s^{(T)}_{i} ={\cal H}(m_{i}^{(T)})=\left \{ 
\begin{array}{lll}
1                &if& m^{(T)}_{i} > 0.5 \\ 
\sigma^{(T)}_{i} &if& m^{(T)}_{i} = 0.5 \\ 
0                &if& m^{(T)}_{i} < 0.5       
\end{array}\equiv
\begin{array}{lll}
 2\omega^{(T)}_{i} > \Omega{(T)} \\ 
 2\omega^{(T)}_{i} = \Omega{(T)} \\ 
 2\omega^{(T)}_{i} < \Omega{(T)} \,.
\end{array}\right .
\end{equation}

\par 
The choice of the memory factor $\alpha$ tunes the memory effect: the limit case
 $\alpha =1$ corresponds to a memory with equally weighted records 
($full$ memory, equivalent to unlimited trailing \textit{majority} memory), whereas $\alpha \ll 1$ intensifies the contribution
 of the most recent states and diminishes the contribution of the more remote states (short-term  
 memory). The choice $\alpha = 0$ leads to the ahistoric model. Due to the rounding \eqref{rounding}\,,
 $\alpha$-memory is not effective if $\alpha \le 0.5$\,.

\par

\section{Actin automata  with memory}

\begin{figure}[!tbp]
 \centering
 \subfigure[]{\includegraphics[width=0.30\textwidth]{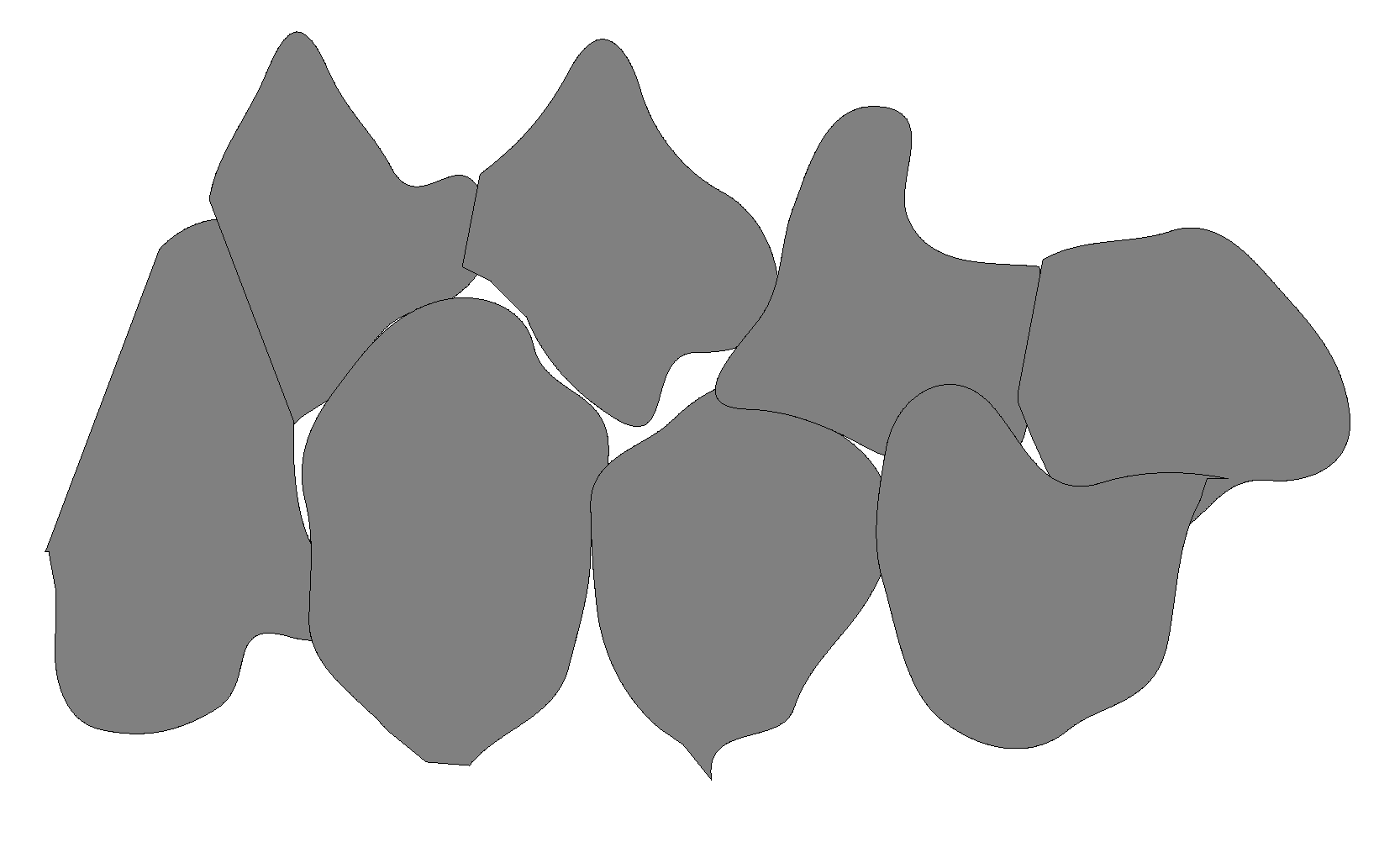}}
 \subfigure[]{\includegraphics[width=0.69\textwidth]{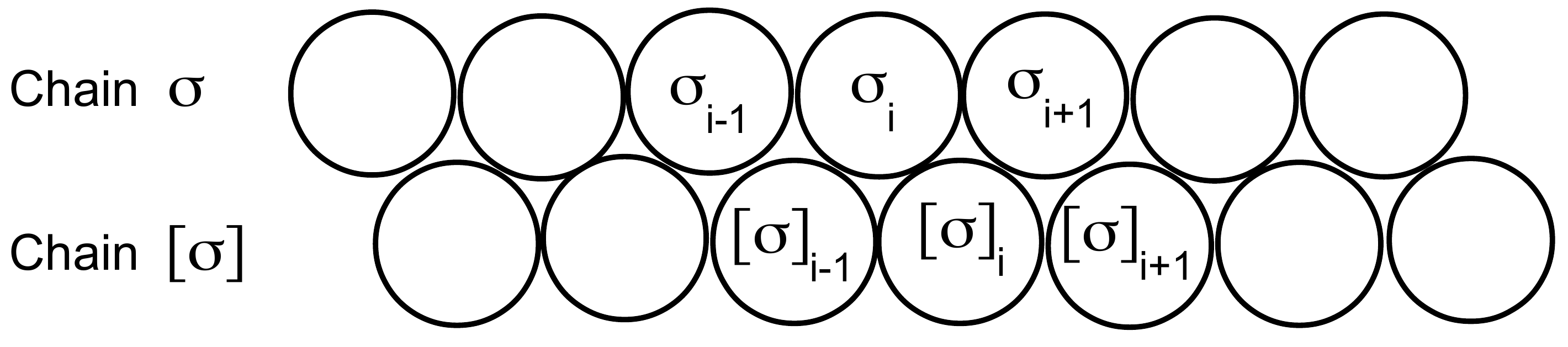}}
\caption{Schematic digram of F-actin strands. (a)~Structure of actin detected by X-ray fiber diffraction. Adapted from~\cite{oda2009nature}. (b)~Actin automata.}
 \label{scheme}
 \end{figure}
 
 Each G-actin molecule (except those at the ends of F-actin strands) has four neighbours, as demonstrated 
 in Fig.~\ref{scheme}. An actin automaton consists of two chains $\sigma$ and $[\sigma]$ of semi-totalistic
  binary-state automata. Each automaton takes two states `0' (resting) and `1' (excited).

\par
Conventional (Markovian) actin CA rules have been proposed to be implemented in two coupled layers (noted $\sigma$ and
$[\sigma]$) with semitotalistic rules~\cite{aadamatzky}\,:

\begin{subequations} \label{actinca}
\begin{align}
\sigma^{(T+1)}_{i} =\left \{ 
\begin{array}{lll}\vspace{0.25cm}
\phi(\sigma^{(T)}_{i-1}+\sigma^{(T)}_{i+1}+[\sigma^{(T)}_{i}]+[\sigma^{(T)}_{i-1}]) &if& \sigma^{(T)}_{i}=0 \\ 
\psi(\sigma^{(T)}_{i-1}+\sigma^{(T)}_{i+1}+[\sigma^{(T)}_{i}]+[\sigma^{(T)}_{i-1}]) &if& \sigma^{(T)}_{i}=1    
\end{array}\right .
\\ \vspace{0.5cm}
[\sigma^{(T+1)}_{i}] =\left \{ 
\begin{array}{lll}\vspace{0.25cm}
\phi([\sigma^{(T)}_{i-1}]+[\sigma^{(T)}_{i+1}]+\sigma^{(T)}_{i}+\sigma^{(T)}_{i+1}) &if& [\sigma^{(T)}_{i}]=0 \\ 
\psi([\sigma^{(T)}_{i-1}]+[\sigma^{(T)}_{i+1}]+\sigma^{(T)}_{i}+\sigma^{(T)}_{i+1}) &if& [\sigma^{(T)}_{i}]=1    
\end{array}\right .
\end{align}
\end{subequations}
\par 
With the subrules $(\phi,\psi)$ expressed in binary and decimal forms as 
$\phi=(\beta_0\beta_1\beta_2\beta_3\beta_4)\equiv \displaystyle {\sum^{4}_{i=0}2^{4-i} \beta_i} $\,, and
$\psi=(\gamma_0\gamma_1\gamma_2\gamma_3\gamma_4)\equiv \displaystyle {\sum^{4}_{i=0}2^{4-i} \gamma_i} $\,.

\par
Actin CA rules with memory will be implemented from   \eqref{actinca} as\,:

\begin{subequations} \label{mactinca}
\begin{align}
\sigma^{(T+1)}_{i} =\left \{ 
\begin{array}{lll}\vspace{0.25cm}
\phi(s^{(T)}_{i-1}+s^{(T)}_{i+1}+[s^{(T)}_{i}]+[s^{(T)}_{i-1}]) &if& s^{(T)}_{i}=0 \\ 
\psi(s^{(T)}_{i-1}+s^{(T)}_{i+1}+[s^{(T)}_{i}]+[s^{(T)}_{i-1}]) &if& s^{(T)}_{i}=1    
\end{array}\right .
\\ \vspace{0.5cm}
[\sigma^{(T+1)}_{i}] =\left \{ 
\begin{array}{lll}\vspace{0.25cm}
\phi([s^{(T)}_{i-1}]+[s^{(T)}_{i+1}]+s^{(T)}_{i}+s^{(T)}_{i+1}) &if& [s^{(T)}_{i}]=0 \\ 
\psi([s^{(T)}_{i-1}]+[s^{(T)}_{i+1}]+s^{(T)}_{i}+s^{(T)}_{i+1}) &if& [s^{(T)}_{i}]=1    
\end{array}\right .
\end{align}
\end{subequations}

\begin{figure}[!tbp]\centering
\includegraphics[width=1.0\textwidth,draft=false]{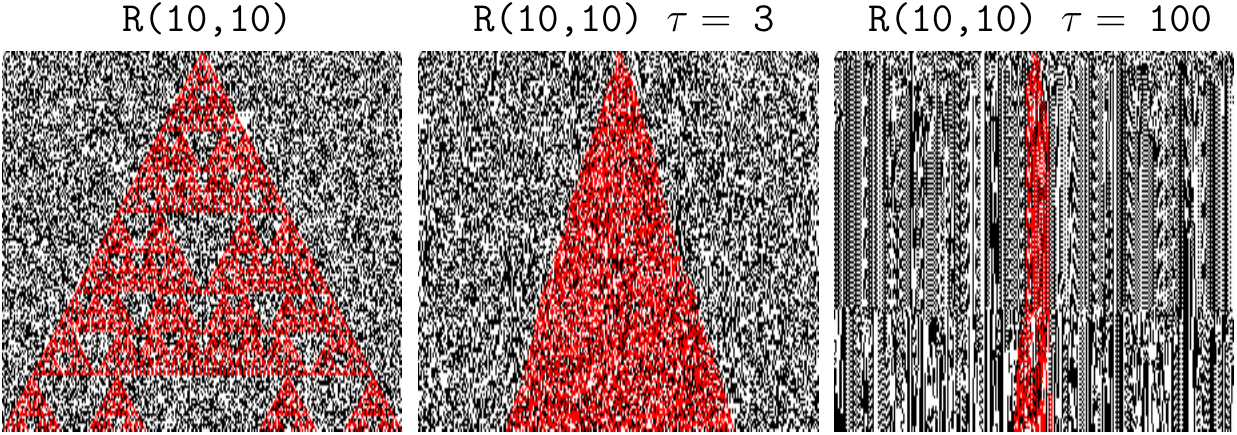}\par\medskip
\includegraphics[width=1.0\textwidth,draft=false]{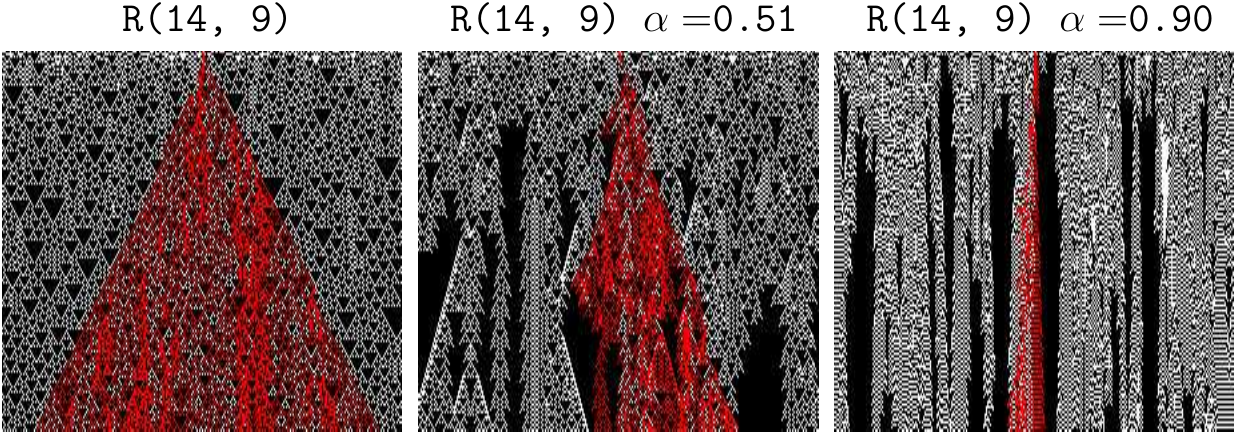}
\caption{Dynamics in two actin rules up to $T=150$\,. Top\,: Rule R(10,10). Ahistoric (left) and  majority memory models.
Bottom\,: Rule R(14,9). Ahistoric (left) and $\alpha$-memory models.}\label{fig:damage}
\end{figure}

Figure \ref{fig:damage} shows the effect of memory  up to $T=150$ in two actin rules
when starting at random, i.e.,  the values of sites are initially uncorrelated and chosen
 at random to be 0 ({\it blank)} or 1 ({\it black}) with probability 1/2 in layers of size $n=300$\,.
 The  pictures  show  also the differences in patterns  resulting  from  reversing the initial center site value. The
{\it perturbed}  region  is enhanced with {\it red} pixels,  corresponding  to  the  site  values  that differed
 among the patterns generated with  the two initial  configurations.
In the top frame of Fig.\,\ref{fig:damage}\,, the actin rule R(10,10)\footnote{R(10,10) is usually referred to as the {\it parity} rule, where both subrules
turn out to be the sum of their inputs modulo two. The simple in form {\it parity} rule has proven to be highly chaotic in CA scenarios.} is shown with ahistoric (left),  
$\tau=3$ and $\tau=100$-majority memory models. 
In the bottom frame, the rule R(14,9) is considered,  with  ahistoric (left),  $\alpha=0.51$ and $\alpha=0.9$ memory models. The inertial (or conservative) effect that memory 
exerts  tends to slow down propagation of growing patterns (perturbations, excitations). This is `slowing down' is manifested by apparent `shrinking of patterns' no the space-time configurations.  This is so even with low memory charge, either in the form $\tau=3$ or as $\alpha=0.51$, but becomes
fully appreciable with high memory (either $\tau=100$, or $\alpha=0.90$). In correspondence, the perturbation spreading in Fig.\,\ref{fig:damage} becomes highly restrained as the
 memory charge increases, so that with high memory charge it remains confined to the proximity of its initial seed.

\par
\begin{figure}[!tbp]\centering
\includegraphics[width=1.0\textwidth,draft=false]{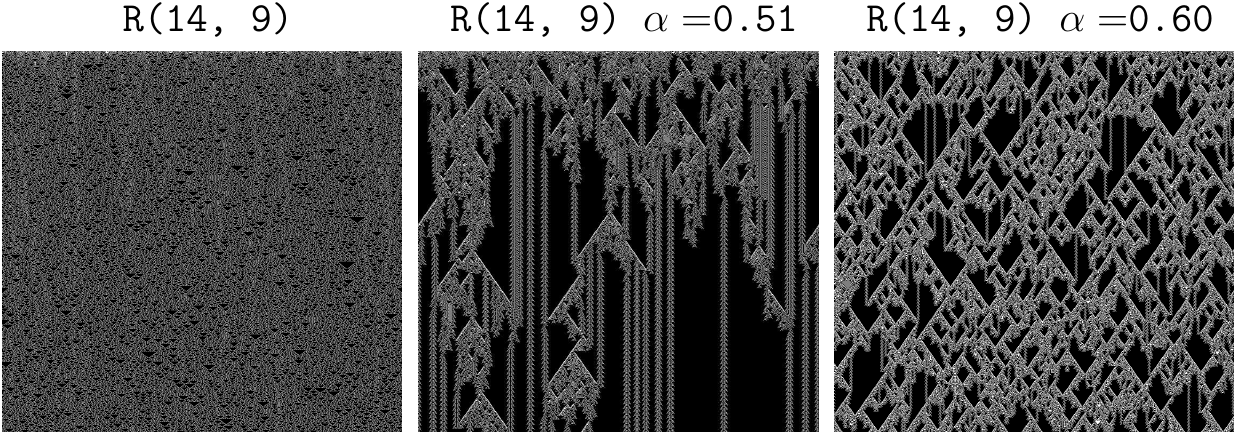}\par\bigskip
\includegraphics[width=1.0\textwidth,draft=false]{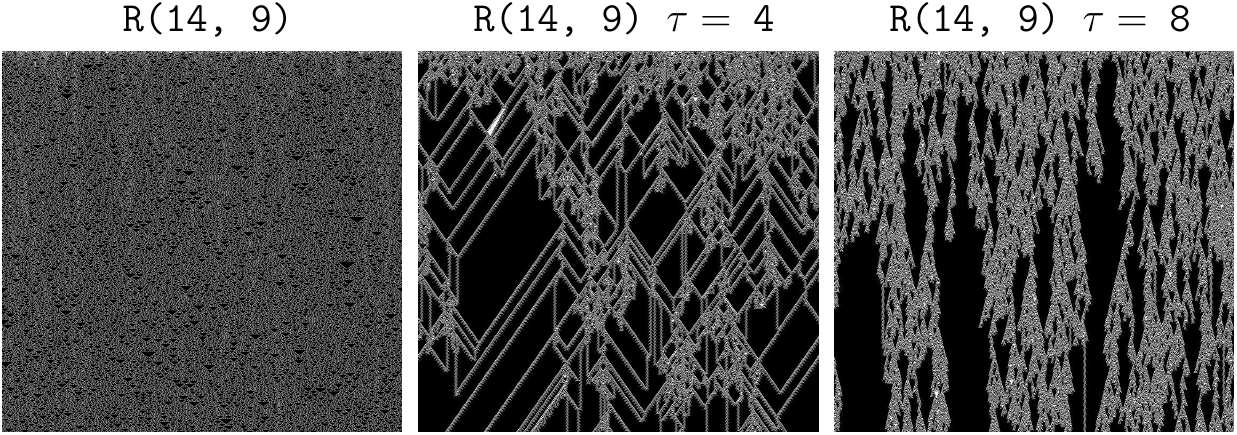}
\caption{Dynamics in the rule R(14,9). Top\,: $\alpha$-memory, Bottom\,: majority memory.}\label{fig:R-14-9}
\end{figure}

Figure \ref{fig:R-14-9} shows the effect of memory  up to $T=1000$ in the actin rule (14,9) when starting at random in layers of size 1500\,.
In the $\alpha$ memory type (top snapshots), even the very low memory charge  $\alpha=0.51$ dramatically alters the conventional (ahistoric) spatio-temporal 
patterns, with  stationary localisations becoming dominating, whereas when $\alpha$ increases to 0.6 a sophisticated spatio-temporal pattern emerges
where travelling glider guns produce gliders, and gliders reflect or annihilate in their collisions. 
With majority memory (bottom snapshots), the low memory charge $\tau=$ induces a {\it monotonous} transition between from enormous amount of gliders generated
 (so many that space is almost completely filled with them), to situation when few stationary glider guns generate gliders with low frequency, 
gliders collide and in most cases annihilate. When $\tau$ reaches 8 only stationary (breathing) localisations persist.

\par
\begin{figure}[!tbp]\centering
\includegraphics[width=1.0\textwidth,draft=false]{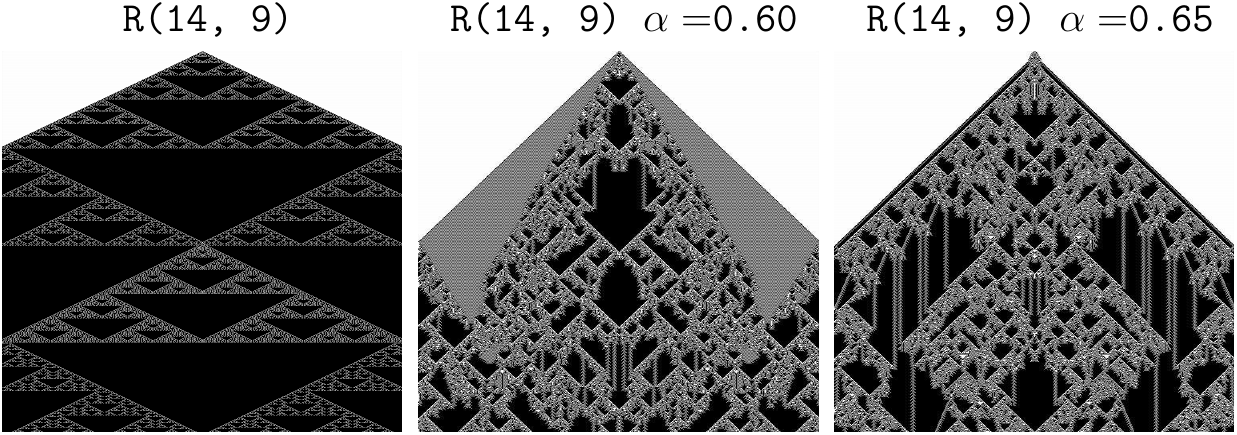}\par\bigskip
\includegraphics[width=1.0\textwidth,draft=false]{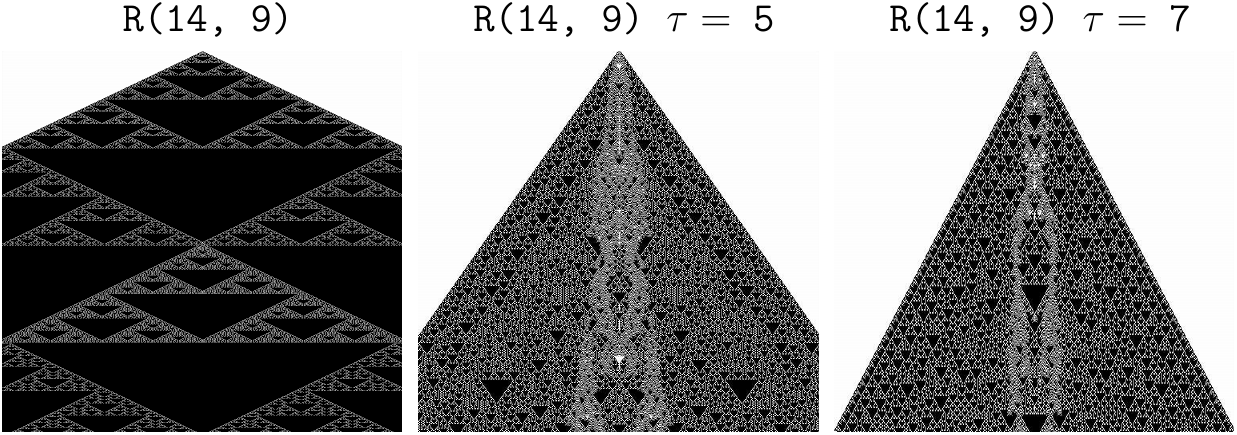}
\caption{Dynamics in the rule R(14,9) starting from one site active seed. Top\,: $\alpha$-memory, Bottom\,: majority memory.}\label{fig:R-14-9-oness}
\end{figure}

 
The richness of the effect of memory may be envisaged in Fig\,.\ref{fig:R-14-9-oness}\,, where rule R(14,9) starts from one site active seed. Sophisticated 
patterns emerge bith the $\alpha$ and majority scenarios shown in   Fig\,.\ref{fig:R-14-9-oness}\,.

\section{Global characteristics}
\par 
Following the reference \cite{aadamatzky}, the following integral measures are calculated on the spatio-temporal pattern $M=(m_{ti})$, where $m_{ti}$ is the state of the 
cell $i$ at time-step $t$\,:
\begin{itemize}
\item Shannon entropy\,: $H=-\displaystyle\sum_{w \in W} \nu(w)/\eta \ln(\nu(w)/\eta)$
\par Where $\eta$ is the sum of the $W$ possible kinds of $3\times3$ configurations found in $M$\,, and $\nu(w)$ is the number of times that the configuration $w$ is found in $M$\,.
\item Simpson diversity\,: $D=1-\displaystyle \sum_{w \in W}
 (\nu(w)/\eta)^2$
\end{itemize}

\par
\begin{figure}[!tbp]\centering
\includegraphics[width=1.0\textwidth,draft=false]{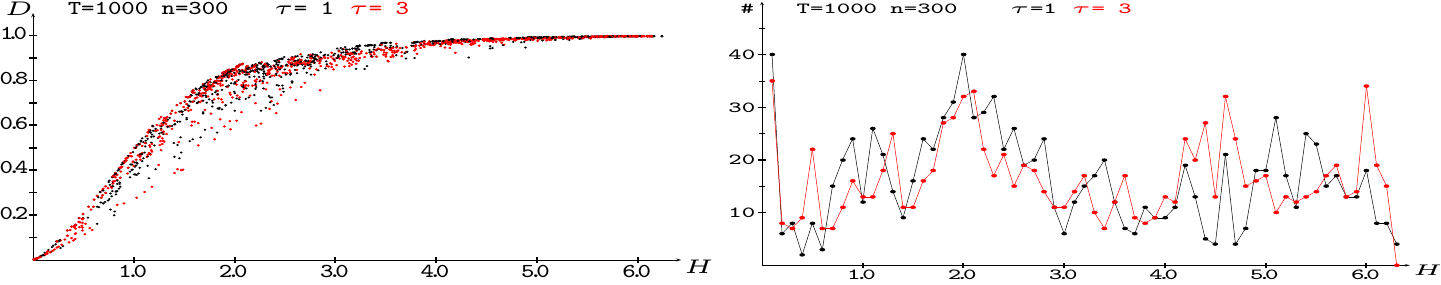}\par\bigskip
\includegraphics[width=1.0\textwidth,draft=false]{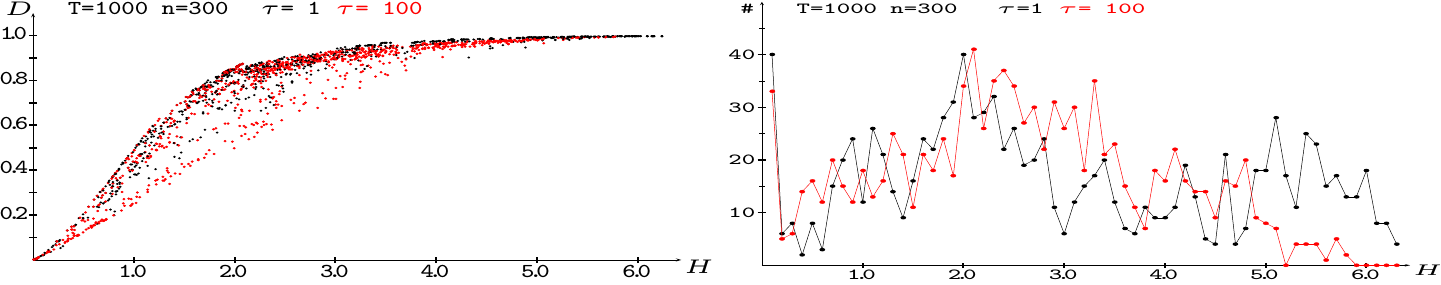}\par\bigskip
\includegraphics[width=1.0\textwidth,draft=false]{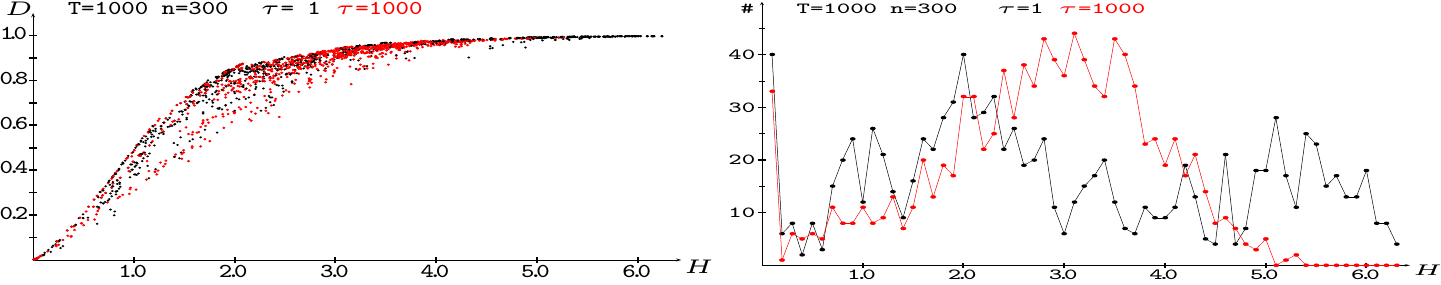}\par\bigskip
\includegraphics[width=1.0\textwidth,draft=false]{entrodivernumber-tau1-1000.pdf}
\caption{Entropy ($H$) versus diversity ($D$) (left) and entropy ($H$) versus number of rules ($\#$) (right) plots. 
Ahistoric (black) and $\tau$- majority (red) memory models. Top\,: $\tau=3$; Middle\,: $\tau=100$; Bottom\,: $\tau=1000$\,. }\label{fig:H-D-Number}
\end{figure}

Figure \ref{fig:H-D-Number} shows the entropy ($H$) versus diversity ($D$) in the left frames, and  the distribution of the number of rules ($\#$) of given entropy.
 Ahistoric (black-marked) and $\tau$-majority memory (red-marked) models are in runs up to $T=1000$ are considered in Fig.\,\ref{fig:H-D-Number}\,: Top\,: $\tau=3$, 
Middle\,: $\tau=100$, Bottom\,: $\tau=1000$\,. Memory seems not to dramatically alter the form of the $H$-$D$ plots, at least at a first visual glance in the left panels of 
Fig.\,\ref{fig:H-D-Number}\,, where the data from the ahistoric and memory simulations appear rather masked. At variance with this, memory appears to notably modify the distribution 
of the number of rules corresponding to the different levels of entropy in the
right panels. Thus, comparing the ahistoric (or even the low level $\tau=3$ memory charge) to the full memory implementation in the bottom-right panel, it becomes apparent that
the highest levels of entropy, close to 6.0, become with no rules filling them in the full memory implementation, whereas most of the rules correspond to the middle level 
H-interval [2.5,3.5]\,, an interval fairly low represented in the ahistoric scenario. Let us take the example of rule R(10,10). In the ahistoric model, R(10,10) achieves
the entropy $H=6.233$, very close to the maximum attainable $H^*= \ln 2^9 =6.238$\,. With $\tau=3$ majority memory the entropy of R(10,10) is lowered to $H=6.005$, and with
full memory to $H=5.335$\,. In the scenarios of Fig.\,\ref{fig:H-D-Number}\, the actin rule R(14,9)  achieves in ahistoric model the entropy 3.342, which atypically increases, 
albeit very little, to 3.835 with $\tau=3$, then decreases to 3.199 with full memory.

\section{Localizations}

We provide additional section on localisations in acting automata with memory for the following reasons. 
Actin filaments are polyelectrolytes surrounded by counter-ions,  the filaments  therefore possess the capacity of transmitting signals or sustaining ionic conductances~\cite{cantiello1991osmotically, tuszynski2004ionic}.  Actin filaments, being rod-like polymers, are particularly likely to have counter-ions adsorbed to their surface  at high ionic concentrations, such as those in the intra-neuronal environment, ions would be expected to densely adsorb to the surface of actin filaments due to complementary charges~\cite{priel2010neural}. The actin filaments are also capable for supporting propagation of discrete breathers (non-linear localised modes of excitation) as a  consequence of nonlinearity in pure, translationally invariant systems of any dimensionality (similar to intrinsic localized modes in anharmonic crystals)~\cite{flach1998discrete,kavitha2013nano, kovaleva2012analytical}. The solitonic signals propagating on actin networks are capable of realising collision-based logical circuits~\cite{adamatzky2001computing, badamatzky, adamatzky2004collision}.

We consider three types of localisations: 
\begin{itemize}
\item glider: travelling localisation, analogues to voltage solitons in cable equation model of actin~\cite{tuszynski2004ionic}, and discrete travelling breathers~\cite{flach1998discrete}. 
\item still life: still stationary patterns, which repeats itself in every time, analogues to a standing wave, or localised excitation in vibrating granular material~\cite{umbanhowar1996localized}
\item oscillator: oscillating stationary localisation, a pattern which repeats itself in a finite number of evolution steps; a glider also repeated itself but it is not stationary), stationary breather~\cite{flach1998discrete} or immobile voltage soliton~\cite{tuszynski2004ionic}.
\end{itemize}
Glider and still life are oscillators as well: a glider is translating oscillator and a still life is stationary oscillator period 1. We compare localisations generated by seeds of five scells in automata without memory and with memory.  Given a rule $\mathcal R$ we say `glider becomes oscillator' if an automaton governed by rule $\mathcal R$ evolves seed $s$  into a glider when it does not have memory and the automaton evolves the seed $s$ into a breather when the rule  $\mathcal R$ is enriched with memory.

\begin{figure}[!tbp]\centering
\includegraphics[width=0.7\textwidth,draft=false]{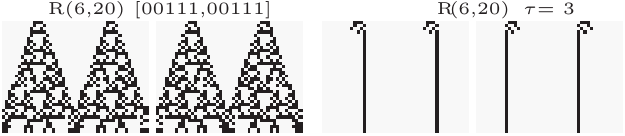}
\caption{Spreading pattern is localised into a still life. Transformation of localisations by majority memory. In each subfigure we see space-time configurations of both actin chains in automata without memory (left) and with memory (right).
 Times goes down.  Rules are indicated above the space-time configurations. Seeds are indicated in the label of the left (ahistoric) simulation.}
\label{stilllife1}
\end{figure}

As we discussed before, introduction of memory typically `shrinks'  patterns. In some rules, e.g. $R(6,20)$ the spreading pattern is shrunk to a still life occupying just one cell (Fig.~\ref{stilllife1}).

\begin{figure}[!tbp]\centering
\subfigure[]{\includegraphics[width=0.70\textwidth,draft=false]{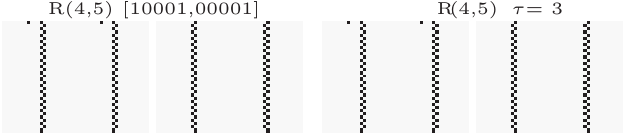}}
\subfigure[]{\includegraphics[width=0.70\textwidth,draft=false]{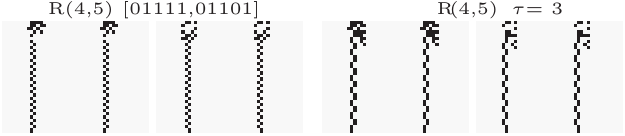}}
\subfigure[]{\includegraphics[width=0.70\textwidth,draft=false]{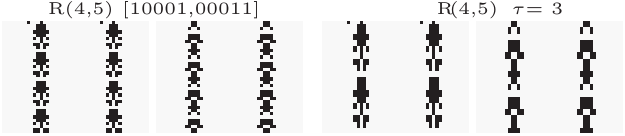}}
\subfigure[]{\includegraphics[width=0.70\textwidth,draft=false]{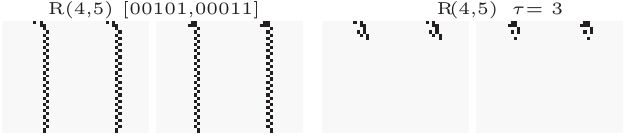}}
\subfigure[]{\includegraphics[width=0.70\textwidth,draft=false]{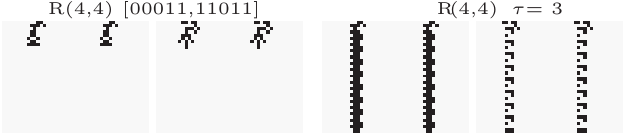}}
\subfigure[]{\includegraphics[width=0.70\textwidth,draft=false]{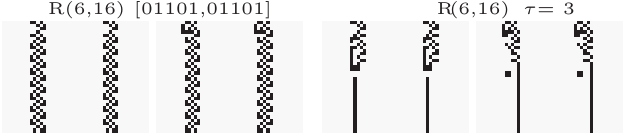}}
\caption{Transformation of oscillators by $\tau=3$ majority memory. In each subfigure we see space-time configurations of both actin chains in automata without memory (left) and with memory (right). Times goes down.  Rules are indicated above the space-time configurations. 
(a)~Oscillator is preserved.
(b )~Oscillator remains yet frequency of its oscillation (breathing) decreases.
(c )~Pattern of oscillations changes. 
(d)~Oscillator is annihilated by memory.
(e)~Oscillator emerges assisted by memory.
(f)~Oscillator is transformed into still life.
Seeds are indicated in the label of the left (ahistoric) simulation.
}
\label{breathers}
\end{figure}

Oscillating stationary localisation is preserved (Fig.~\ref{breathers}a), sometimes with decrease of oscillation frequency  (Fig.~\ref{breathers}b) or 
change of the oscillations pattern (Fig.~\ref{breathers}c) or annihilated  (Fig.~\ref{breathers}d).  In some rules oscillators can emerge, when state-transition rules are enriched with memory, e.g. in  (Fig.~\ref{breathers}d) a seed produces an extinguishing pattern in memoryless automaton yet it produces oscillator when majority memory is introduced.  In rule $R(6,16)$ introduction of memory converts oscillator into a still life. 

\begin{figure}[!tbp]\centering
\subfigure[]{\includegraphics[width=0.8\textwidth,draft=false]{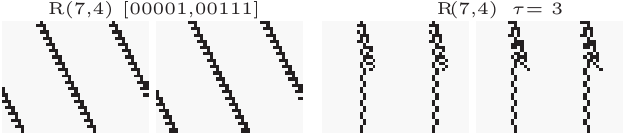}}
\subfigure[]{\includegraphics[width=0.8\textwidth,draft=false]{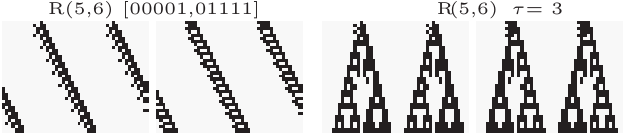}}
\subfigure[]{\includegraphics[width=0.8\textwidth,draft=false]{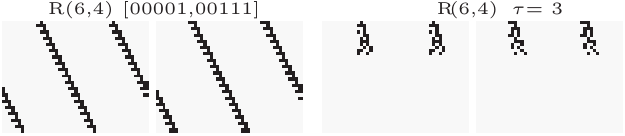}}
\subfigure[]{\includegraphics[width=0.8\textwidth,draft=false]{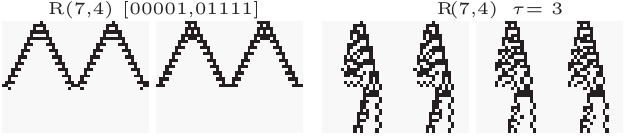}}
\subfigure[]{\includegraphics[width=0.8\textwidth,draft=false]{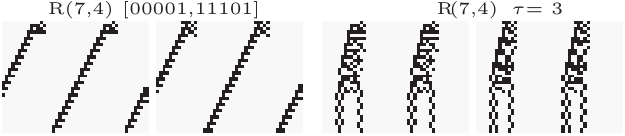}}
 
\caption{Transformation of gliders by  $\tau=3$ majority memory. In each subfigure we see space-time configurations of both actin chains in automata without memory (left) and with memory (right). Times goes down.  Rules are indicated above the space-time configurations. 
(a)~Glider becomes oscillator.
(b)~Glider becomes a propagating and expanding pattern (Sierpinski gasket in this  particular example).
(c )~Glider is annihilated.
(d)~Glider propagating left is transformed into larger slow moving glider and glider propagating right becomes a oscillator.
(e)~Glider `splits' into two breathers.}
\label{gliders}
\end{figure}

Gliders are transformed as follows (Fig.~\ref{gliders}). Glider can be transformed to a oscillator  (Fig.~\ref{gliders}a), or expanding pattern (we can say that the glider was exploded by memory, Fig.~\ref{gliders}b), or annihilated  (Fig.~\ref{gliders}c).  In some cases, where a seed generates two gliders -- one travels left and another travels right -- one of the gliders becomes larger and slow moving glider while another glider is transformed into an oscillator (Fig.~\ref{gliders}d); or, a single gliders is transformed into two oscillators with different patterns of oscillations 
 (Fig.~\ref{gliders}e).

 \begin{figure}[!tbp]\centering
\includegraphics[width=0.98\textwidth,draft=false]{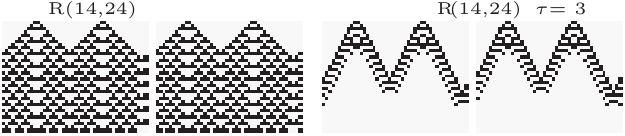}
\caption{Initial configurations not of the 5-seed type. A spreading pattern is trimmed to gliders. Transformation of gliders by majority memory. In each subfigure we see space-time configurations of both actin chains in automata without memory (left) and with $\tau=3$-memory (right). Times goes down.  Rules are indicated above the space-time configurations. Seeds\,:[00110000000000000000000110~,~01100000000000000000001100] 
}
\label{gliders2}
\end{figure}

\begin{figure}[!tbp]\centering
\includegraphics[width=0.98\textwidth,draft=false]{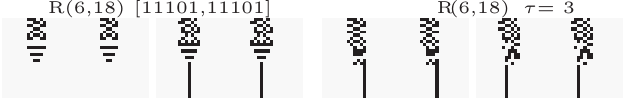}
\caption{Entrainment. Dynamics of actin chains is asynchronous in rules without memory: still life persists only one chain. When memory is introduced both chains host still lifes.
In each subfigure we see space-time configurations of both actin chains in automata without memory (left) and with $\tau=3$-memory (right). Times goes down.  Rules are indicated above the space-time configurations. }
\label{entrainemnt}
\end{figure}

In rare cases, e.g. rule $R(14,24)$  a spreading pattern is converted to two gliders, propagating to the left and to the right. The gliders annihilate when collide (Fig.~\ref{gliders2}). We can also observe an entrainment phenomenon 
(Fig.~\ref{entrainemnt}) where a localisation `inhabiting' only one chain in a rule without memory, spreads to the second chain when memory is introduced. 
 
 \begin{figure}[!tbp]\centering
\includegraphics[width=0.5\textwidth,draft=false]{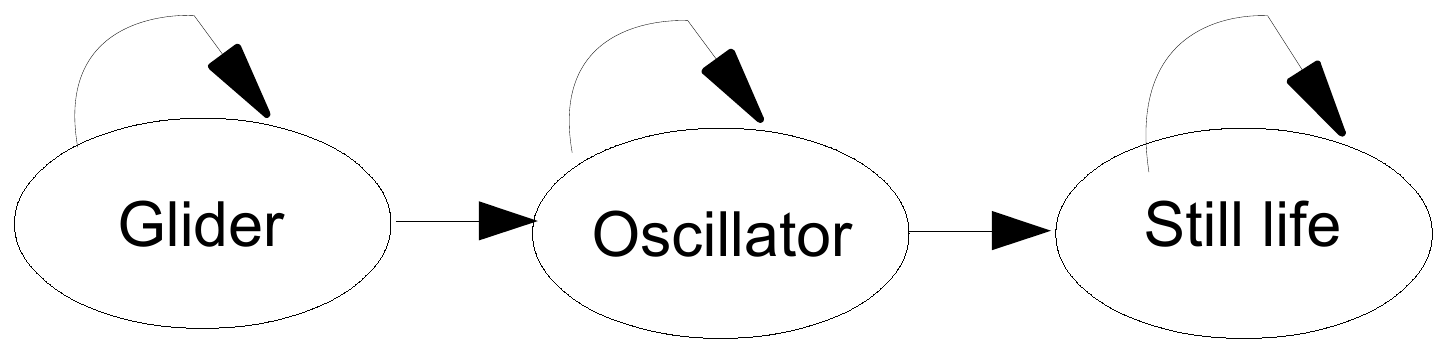}
\caption{Memory-induced transformations of localisation.}
\label{schemetransitions}
\end{figure}

By undertaking exhaustive analysis of the rules we found that only the following transformations of localisations are possible when state transition rules are enriched with memory (Fig.~\ref{schemetransitions}). A glider is transformed to  an oscillator. An oscillator is transformed to a still life. In some cases memory preserved gliders, oscillators and still life. Transformations of still lifes to oscillators or glider, and oscillations to gliders have not been observed so far.  Notably, in elementary cellular automata with memory transitions from glider to oscillation to still life, induced by memory, could be found~\cite{DBLP:journals/fuin/MartinezAA15, martinez2010make, martinez2012complex}. We did not such transitions in actin automata with memory, this could be due to the fact interaction between two one-dimensional automata forming acting filament leads to `inhibition' of glider dynamics and prevents formation of mobile localisations when memory is introduced to node state-transition rules.


\section{Discussion}

\begin{figure}[!tbp]
 \centering
\subfigure[]{\includegraphics[width=0.5\textwidth]{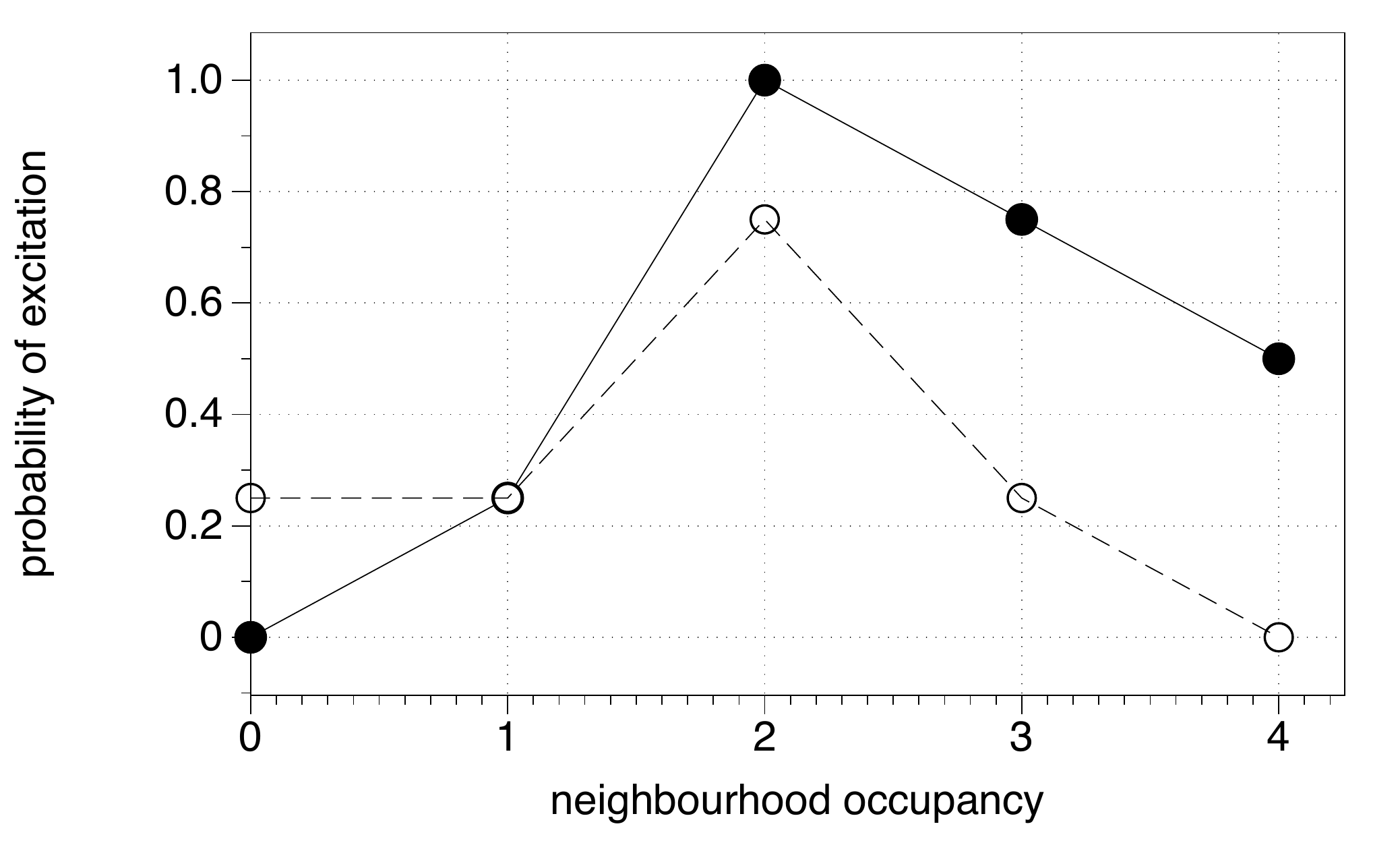}}
\subfigure[]{\includegraphics[width=0.5\textwidth]{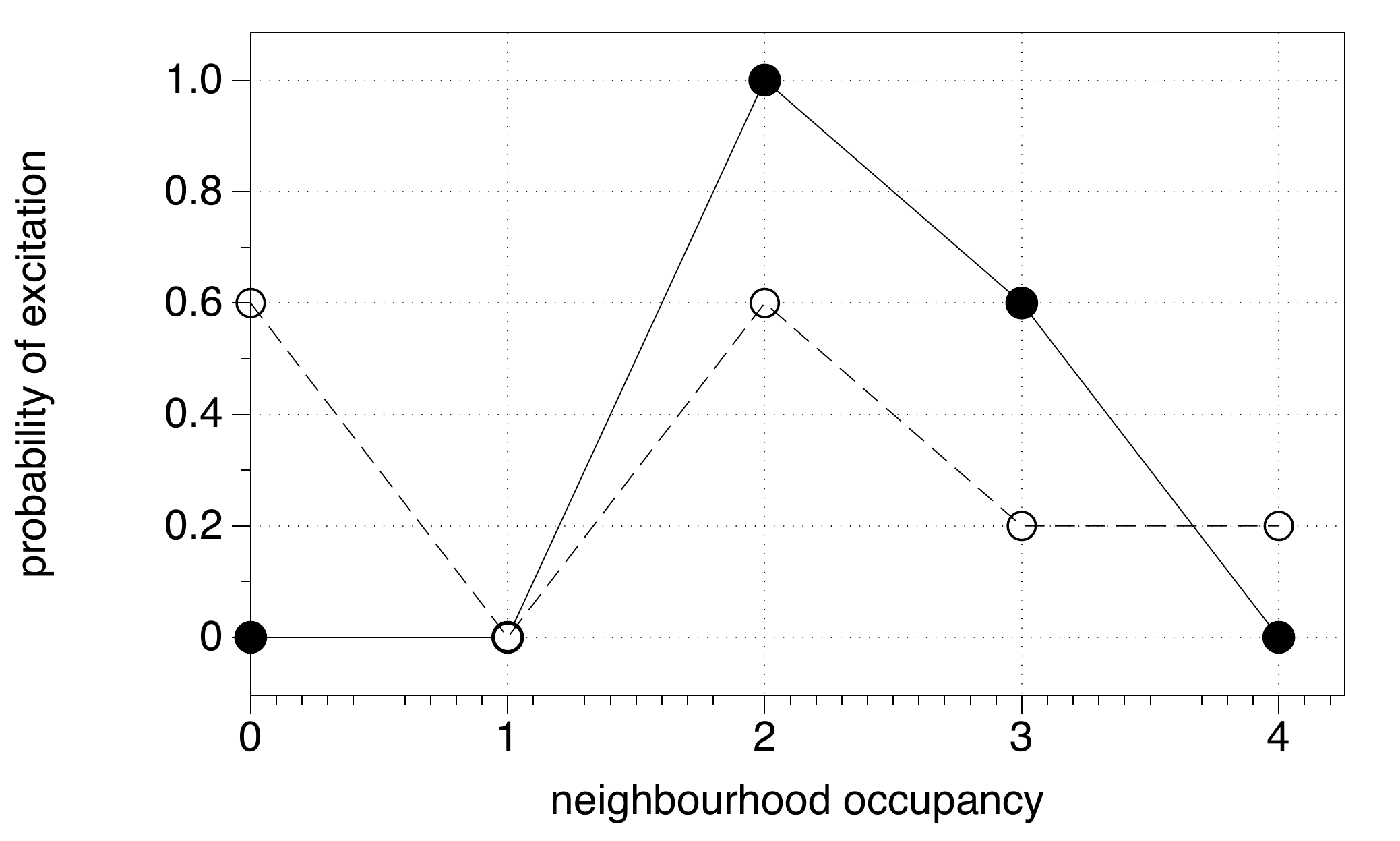}}
  \caption{Dependence of a probability of excitation of a node in actin automaton (a)~supporting travelling 
  localizations and (b)~supporting stationary localizations  on a number of excited neighbours of the node. The plots are calculated for rules R(7,4), R(5,6), R(6,4), R(12,24), which support travelling localisations shown in Figs.~\ref{gliders}, 
\ref{gliders2}, \ref{fig:R-7-4-interactions}, and rules R(6,20), R(4,5), R(4,4), R(6,16), R(6,18), which support stationary localisations shown in Fig.~\ref{stilllife1}, \ref{breathers}, \ref{entrainemnt}.    Solid discs, connected by solid line, show probability of excitation of a resting node, circles,  connected by dashed line, show probability of excitation of an excited node, i.e. of an excited node to remain excited.}
 \label{travellingpolynomial}
 \end{figure}
 
In our original studies of actin automata (without memory)~\cite{adamatzky2014actin} we demonstrated, by exhaustive search of localisations over the whole rule space, that rule supports travelling localizations if a resting node excites if it has two or four excited neighbours  and an excited node remains excited if it has no excited neighbours (there are indeed additional modes of excitability necessary to support propagation of the gliders). We also found that a rule supports stationary localizations if a resting node excites if it has three excited neighbours  and an excited node remains excited if it has less than four excited neighbours. Analysing the rules of actin automat with majority, which we used to illustrate transformation of travelling and stationary localisations, we found that two excited neighbours is a necessary prerequisite for a resting node to be excited and for an excited node to remain excited for a rule to support travelling (Fig.~\ref{travellingpolynomial}a) and stationary (Fig.~\ref{travellingpolynomial}b) localisations. Second most common occupancy for these rules is three excited neighbours.

\begin{figure}[!tbp]\centering
\subfigure[]{\includegraphics[width=0.47\textwidth,draft=false]{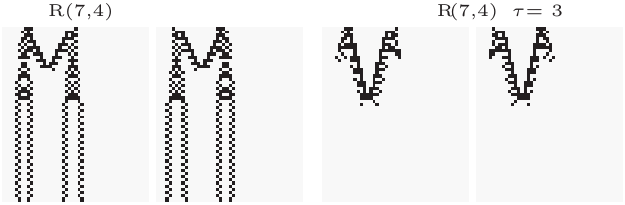}} \subfigure[]{\includegraphics[width=0.47\textwidth,draft=false]{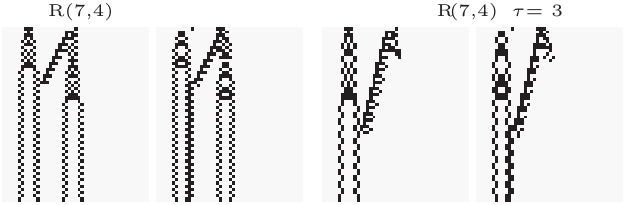}}
\caption{Initial configurations not of the 5-seed type. Interactions of localizations of rule R(7,4) with (right) and without (left) $\tau=3$ majority memory. (a)~Two gliders collide and annihilate in the result of the collision. (b)~Glider is stopped by oscillator. In each subfigure we see space-time configurations of both actin chains in automata. Seeds\,:
(a) [0011100000000000001,0100000000000001110]\, (b) [11100000000000001,00000010000000111]}
\label{fig:R-7-4-interactions}
\end{figure}

By enriching actin automata with memory we demonstrated that when local transitions rules are enriched with memory they generate slower propagation or expanding patterns, generate less `complex', as measured by entropies, space-time configurations. The memory enrichment also leads to transformation of gliders to oscillatory stationary localisations to still stationary localisations.  These  results well complement our previous findings in  abstract analysis of intracellular acting filaments and complement parallel developments, such as extensions of actin automata to  quantum actin automata~\cite{siccardi2015actin} and cable equations models of voltage solitons in actin~\cite{siccardi2015boolean}.  Gliders are propagating signals. Stationary localisation are memory --- in a sense of volatile computer memory.  So by varying depth of actin units memory we can force actin filaments either to produce signals --- which compute when collide --- or to act as a memory device.   Basic computing properties of the quantum actin automata --- Boolean logic gates and binary adder~\cite{siccardi2015actin}, and three valued logic operations~\cite{cadamatzky} --- have not yet been matched in the actin automata memory. These will topics of our further studies.

 Right now we can just demonstrate interactions which can be interpreted in terms of collision-based computing: presence of a localisation indicate logical {\sc True} and absence --- logical {\sc False}.   A `classical' annihilation gate $\langle x, y \rangle \rightarrow \langle  \overline{x} y,  x \overline{y}  \rangle$ is shown in Fig.~\ref{fig:R-7-4-interactions}a for inputs $x=${\sc True} and $y=${\sc False}.  When  $x=${\sc True} there is a glider travelling right, when  $y=${\sc True} there is a glider travelling left. If just one glider present in the gate, it continue it movement undisturbed. The trajectory of undisturbed movement of the glider travelling right represents $x \overline{y}$ and the trajectory of undisturbed movement of the glider travelling left represents $\overline{x} y$. Travelling localisations, representing data and result of computation can be stopped via stationary oscillators. An example is shown in Fig.~\ref{fig:R-7-4-interactions}b. Glider travels left. It collides into a oscillator. The oscillator changes its states in the result of collision. Thus travelling data are re-written into stationary data.

 \begin{figure}[!tbp]\centering
\subfigure[]{\includegraphics[width=0.47\textwidth,draft=false]{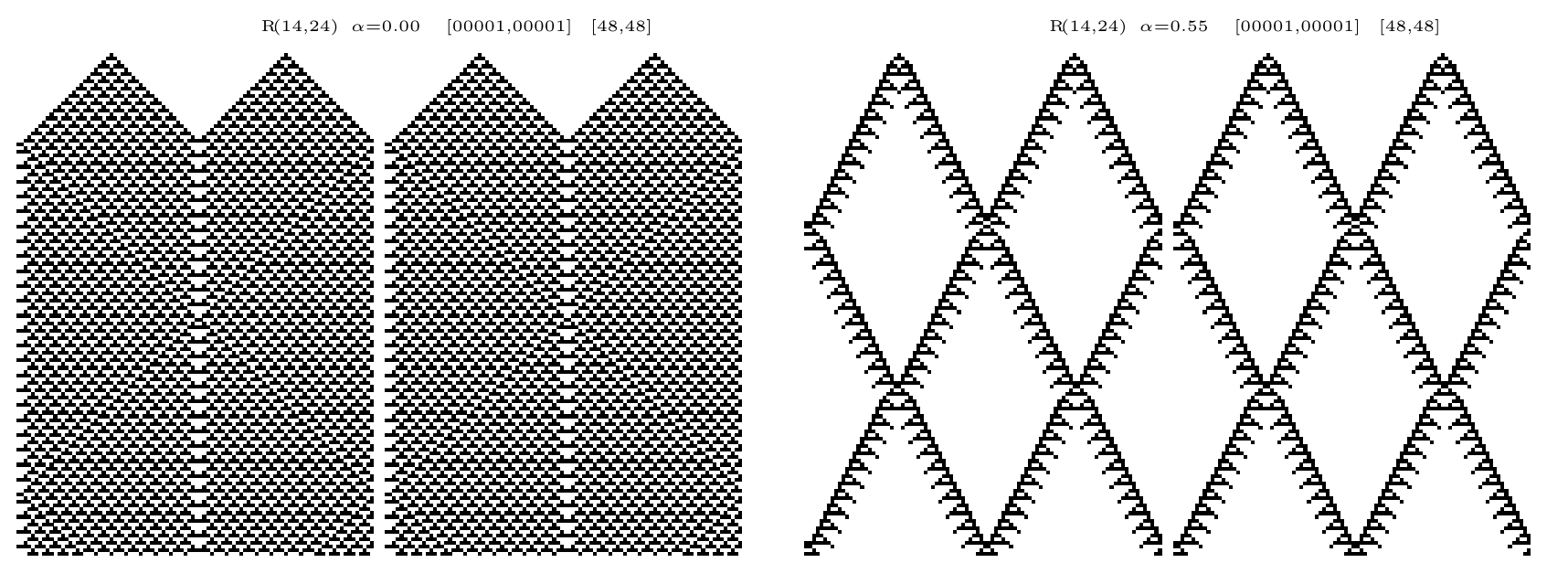}} 
\subfigure[]{\includegraphics[width=0.47\textwidth,draft=false]{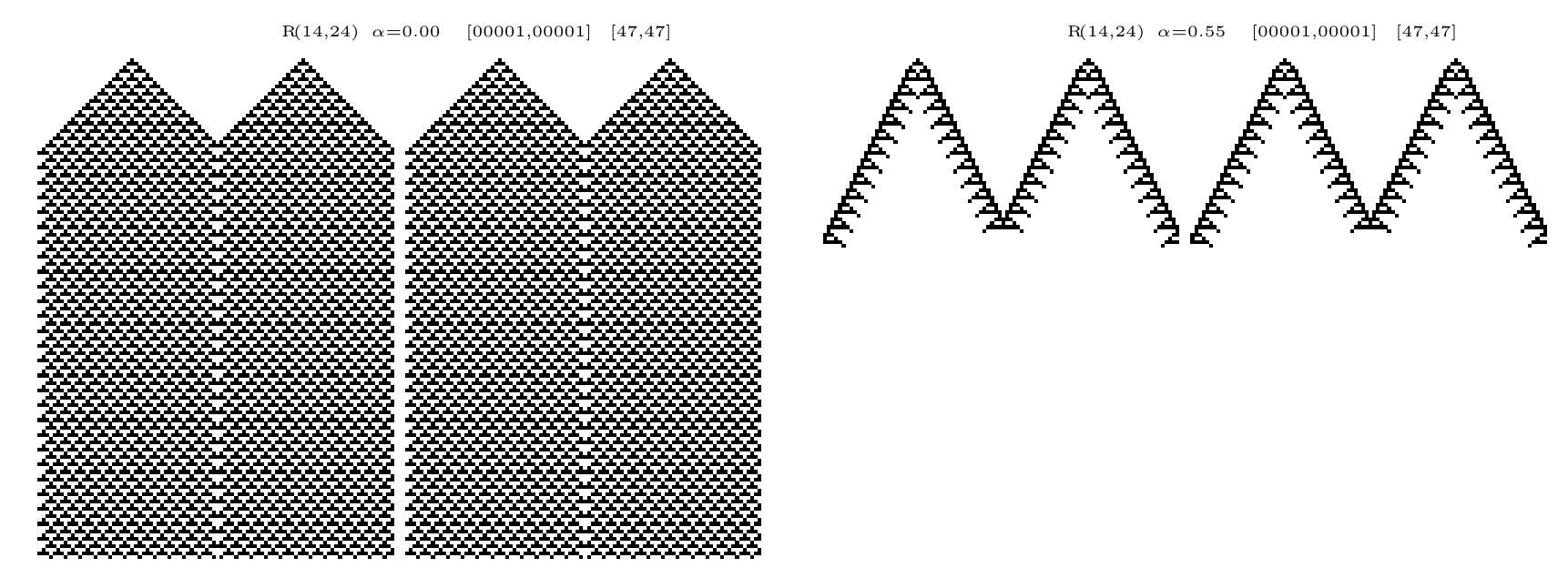}}
\caption{Emergence of localisation induced by memory, rule R(12,24). Distance between glider seeds is
(a)~48 nodes and (b)~51 nodes.  In each subfigure we see space-time configurations of both actin chains in automata without memory (left) and with $\alpha=0.55$-memory (right). Initially just single node is in state `1'.
Times goes down.  Rules are indicated above the space-time configurations.}
\label{fig:R14-24-alpha0-055}
\end{figure}

Another example of memory-induced emergence of collision-based gates is illustrated in Fig.~\ref{fig:R14-24-alpha0-055}. In actin automata rule R(14,24) seeds of a single node in state `1' develop into propagating patterns of regular `excitations' of the actin filaments. When memory is introduced the interior of the propagating patterns is `filtered out' and only wave-fronts, i.e. gliders, remain. The gliders reflect and change their phases when a distance between seeds is 48 nodes (Fig.~\ref{fig:R14-24-alpha0-055}a) and the gliders annihilate when the distance between seeds is 51 nodes (Fig.~\ref{fig:R14-24-alpha0-055}b).

\section*{Acknowledgement}

We acknowledge the financial support of the Future and Emerging Technologies (FET) programme within the Seventh Framework Programme for Research of the European Commission, under the Collaborative project PhyChip, grant agreement number 316366.

\bibliographystyle{ws-ijbc}
\bibliography{bibmactin}

\end{document}